\documentclass{article}
\usepackage{spconf,amsmath,graphicx}

\usepackage{amsmath,graphicx}
\usepackage[T1]{fontenc}
\usepackage{amsfonts}
\usepackage{amssymb}
\usepackage{bm}
\usepackage[noadjust]{cite}
\usepackage[nolist,printonlyused]{acronym}
\usepackage{algorithm}
\usepackage{algpseudocode}
\usepackage{mathtools}

\DeclarePairedDelimiter\ceil{\lceil}{\rceil}

\newcommand\eqbydef{\stackrel{\mathclap{\normalfont\mbox{\tiny def}}}{=}}

\DeclareFontFamily{U}{mathx}{\hyphenchar\font45}
\DeclareFontShape{U}{mathx}{m}{n}{
	<5> <6> <7> <8> <9> <10>
	<10.95> <12> <14.4> <17.28> <20.74> <24.88>
	mathx10
}{}
\DeclareSymbolFont{mathx}{U}{mathx}{m}{n}
\DeclareFontSubstitution{U}{mathx}{m}{n}
\DeclareMathAccent{\widebar}{0}{mathx}{"73}

\usepackage{tikz}
\usepackage{pgfplots}
\pgfplotsset{compat=newest}
\usetikzlibrary{spy}

\def\figsize{0.66\textwidth}

\usepackage{times}

\usepackage{glossaries}
\newacronym{ris}{RIS}{Reconfigurable intelligent surface}

\title{Double-RIS Versus Single-RIS Aided Systems: Tensor-Based MIMO Channel Estimation and Design Perspectives}
\name{Khaled Ardah, Sepideh Gherekhloo, André L. F. de Almeida, Martin Haardt\thanks{The authors gratefully acknowledge the support of the German Research Foundation (DFG) under contract no.~HA 2239/6-2 (EXPRESS II) and the support of CAPES/PRINT (Grant no. 88887.311965/2018-00). The research of André L. F. de Almeida is partially supported by the CNPq (Grant no. 306616/2016-5).}  \thanks{K. Ardah, S. Gherekhloo,, and M. Haardt are with the Communications Research Laboratory (CRL), TU Ilmenau, Ilmenau, Germany (e-mail: \{khaled.ardah, sepideh.gherekhloo, martin.haardt\}@tu-ilmenau.de). A. L. F. de Almeida is with the Wireless Telecom Research Group (GTEL), Federal University of Cear\'a, Fortaleza, Brazil (e-mail: andre@gtel.ufc.br).}  }
\address{}
\def\ka#1{\textcolor{black}{#1}}
\def\kr#1{\textcolor{black}{#1}}

\begin{document}
%\ninept
%
\maketitle
\begin{abstract}
Reconfigurable intelligent surfaces (RISs) have been proposed recently as new technology to tune the wireless propagation channels in real-time. However, most of the current works assume single-RIS (S-RIS)-aided systems, which can be limited in some application scenarios where a transmitter might need a multi-RIS-aided channel to communicate with a receiver. In this paper, we consider a double-RIS (D-RIS)-aided MIMO system and propose an alternating least-squared-based channel estimation method by exploiting the Tucker2 tensor structure of the received signals. Using the proposed method, the cascaded MIMO channel parts can be estimated separately, up to trivial scaling factors. Compared with the S-RIS systems, we show that if the RIS elements of a S-RIS system are distributed carefully between the two RISs in a D-RIS system, the training overhead can be reduced and the estimation accuracy can also be increased. Therefore, D-RIS systems can be seen as an appealing approach to further increase the coverage, capacity, and efficiency of future wireless networks compared to S-RIS systems.
\end{abstract}
\vspace{-5pt}
\section{Introduction}\label{sec:intro}
\vspace{-8pt}
\glspl{ris} have been proposed recently as {a} cost-effective {technology} for reconfiguring the propagation channels in wireless communication systems \cite{comMagazine}. An RIS is a 2D surface equipped with a large number of tunable units that can be realized using, e.g., inexpensive antennas or metamaterials and controlled in real-time to influence the communication channels without generating its own signals. Recently, \gls{ris}-aided communications have attracted great attention \cite{irs}, due to their potential of improving the efficiency, the communication range, and the capacity of wireless communication systems.

Most of the current works, e.g., in \cite{RISCapacity,ardah2020trice,gherekhloo2021tensor,BF1,andre,andreRIS}, assume single-RIS (S-RIS)-aided systems, where a transmitter (Tx) communicates with one receiver (Rx), or more, via a single RIS-aided channel. However, in many application scenarios, e.g., in a urban area or in a satellite-to-indoor communication, the Tx might need a multi-RIS-aided channel to have successful communication with the Rx. Moreover, in a S-RIS system, it was shown that the RIS should be either deployed closer to the Tx or closer to the Rx to achieve the best performance gain \cite{Emil}. This fundamental result gives rise to the double-RIS (D-RIS) systems, where one RIS is deployed closer to the Tx and another is deployed closer to the Rx. In such systems, channel estimation (CE) becomes more problematic since the cascaded (effective) channel contains three parts not only two as in the S-RIS systems (see Fig. \ref{fig:fig1}). 

In this paper\footnote{\textbf{Notation:} The conjugate, the transpose, the conjugate transpose (Hermitian), the pseudoinverse, the Kronecker product, and the Khatri-Rao product are denoted as ${\bm A}^{*}$, ${\bm A}^{\mathsf{T}}$, ${\bf A}^{\mathsf{H}}$, ${\bf A}^{+}$, $\otimes$, and $\diamond$, respectively. Moreover, $\bm{1}_N$ is the all ones vector of length $N$, ${\bf I}_N$ is the $N\times N$ identity matrix, $\text{diag}\{{\bf a}\}$ forms a diagonal matrix ${\bf A}$ by putting the entries of the input vector ${\bf a}$ \kr{on} its main diagonal, $\text{vec}\{ {\bf A} \}$ forms a vector by staking the columns of ${\bf A}$ over each other, $\text{unvec}\{{\bf A}\}$ is the reverse of the vec operator, $\ceil*{x}$ is the ceiling function, and the $n$-mode product of a tensor $\bm{\mathcal{A}}\in \mathbb{C}^{I_1\times I_2\times \dots,\times I_N}$ with a matrix ${\bf B}\in \mathbb{C}^{J\times I_n} $ is denoted as $\bm{\mathcal{A}} \times_n {\bf B}$.. Moreover, the following properties are used: {Property~1}: $\text{vec}\{ {\bf A}  {\bf B} {\bf C} \} = ({\bf C}^{\mathsf{T}} \otimes {\bf A}) \text{vec} \{{\bf B}\}$ and	{Property~2}: $\text{vec}\{ {\bf A} \text{diag}\{ {\bf b}\} {\bf C} \} = ({\bf C}^{\mathsf{T}}\diamond{\bf A}) {\bf b}$.}, we consider a D-RIS aided MIMO system and propose an efficient CE method by exploiting the tensor structure of the received signals \cite{Kolda,andre_tensors,ardah2019wsa}. Specifically, we first show that the received signals in flat-fading D-RIS aided MIMO systems can be arranged in a 3-way tensor that admits a Tucker2 decomposition \cite{Kolda}. Accordingly, an alternating least-squared (ALS)-based method is proposed, where the {Tx}-to-RIS 1 channel (denoted by ${\bf H}_{\text{T}}$), the-RIS~1-to-RIS~2 channel (denoted by ${\bf H}_{\text{S}}$), and the RIS~2-to-{Rx} channel (denoted by ${\bf H}_{\text{R}}$) can be estimated separately, up to trivial scaling factors. We compare the proposed ALS method for D-RIS systems to the ALS method for S-RIS systems proposed in \cite{andreRIS,andre} in terms of the minimum training overhead and the estimation accuracy. {It is shown that if the RIS elements in the S-RIS systems are distributed carefully between the two RISs in the D-RIS systems, the training overhead can be reduced and the estimation accuracy can also be increased.} 
Note that, since the system spectral efficiency is inversely proportional to the length of the training overhead, we conjecture that there is an optimal distribution of the RIS elements that strikes an optimal trade-off between the training overhead and the achievable performance, which is out of the scope of this paper and we leave for future work.  It is worth mentioning that the considered D-RIS system can also resemble communication scenarios where the RISs in both communicating ends are co-located with the transceivers, as it has been proposed in \cite{CoLocated}. Therefore, ${\bf H}_{\text{T}}$ and ${\bf H}_{\text{R}}$ channels can be assumed known by careful transceivers design.

	\vspace{-10pt}
\section{D-RIS System Model}
\begin{figure}[t]
	\centering
	\includegraphics[width=0.49\textwidth]{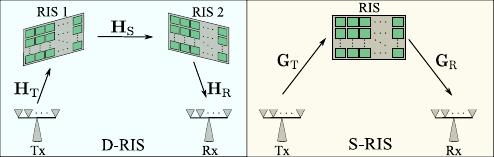}
	%	\vspace{-5pt}
	\caption{{D-RIS versus S-RIS aided MIMO communications.}}
%	\vspace{-15pt}
	\label{fig:fig1}
\end{figure}

	\vspace{-10pt}
In this section, we consider a D-RIS-aided MIMO communication system as depicted \kr{on} the left-side of Fig. \ref{fig:fig1}, where a {Tx} with $M_{\text{T}}$ antennas is communicating with a {Rx} with $M_{\text{R}}$ antennas via a D-RIS-aided channel. Here, RIS 1 is assumed to be close to the {Tx} and has $N_1$ reflecting elements, while RIS 2 is assumed to be close to the {Rx} and has $N_2$ reflecting elements. We assume that the {Tx}-to-{Rx}, the {Tx}-to-RIS 2, and the RIS~1-to-{Rx} channels are unavailable due to blockage or too weak due to high pathloss.

Let ${\bf H}_{\text{T}} \in \mathbb{C}^{N_{1}\times M_{\text{T}}}$ be the {Tx}-to-RIS 1 channel, ${\bf H}_{\text{S}} \in \mathbb{C}^{N_{2}\times N_{1}}$ be the RIS~1-to-RIS 2 channel, and ${\bf H}_{\text{R}} \in \mathbb{C}^{M_{\text{R}}\times N_{2}}$ be the RIS 2-to-{Rx} channel. To estimate these channels, we conduct a channel-training procedure, which occupies $ L = I \cdot K$ subframes. The received signal at the $(i,k)$th subframe, $i \in \{1,\dots,I\}$ and $k \in \{1,\dots,K\}$, can be written as 
\begin{align}
	{\bf \bar y}_{i,k} = {\bf H}_{\text{R}} \bm{\Phi}_i {\bf H}_{\text{S}} \bm{\Psi}_i {\bf H}_{\text{T}} {\bf  f}_{k} s_{k} + {\bf \bar n}_{i,k} \in \mathbb{C}^{M_\text{R}},
\end{align}
where $\bm{\Psi}_i = \text{diag}\{ \bm{\psi}_i \} \in \mathbb{C}^{N_1 \times N_1}$ is the $i$th diagonal reflection matrix of RIS 1, with $\bm{\psi}_i \in \mathbb{C}^{N_1}$ and $|[\bm{\psi}_i]_{[j]}| = 1/\sqrt{N_1}$, $\bm{\Phi}_i = \text{diag}\{ \bm{\phi}_i \} \in \mathbb{C}^{N_2 \times N_2}$ is the $i$th diagonal reflection matrix of RIS 2, with $\bm{\phi}_i \in \mathbb{C}^{N_2}$ and $|[\bm{\phi}_i]_{[j]}| = 1/\sqrt{N_2}$, ${\bf  f}_{k} \in \mathbb{C}^{M_\text{T}}$ is the $k$th precoding vector at the {Tx} with $\Vert {\bf  f}_{k} \Vert = 1$, $s_k$ is the $k$th unit-norm training symbol, and ${\bf \bar n}_{i,k} \in \mathbb{C}^{M_{\text{R}}}$ is the additive white Gaussian noise vector {having zero-mean circularly symmetric complex-valued entries} with variance $\sigma^2$.

Let $\bm{\Psi} = [\bm{\psi}_1, \dots, \bm{\psi}_I] \in \mathbb{C}^{N_1 \times I}$, $\bm{\Phi} = [\bm{\phi}_1, \dots, \bm{\phi}_I] \in \mathbb{C}^{N_2 \times I}$, and ${\bf F} = [{\bf  f}_{1} s_{1}, \dots, {\bf  f}_{K} s_{K}] \in \mathbb{C}^{M_\text{T} \times K}$. Then, by staking $\{{\bf \bar y}_{i,k}\}_{k = 1}^{K}$ next to each other as $	{\bf \bar Y}_{i} = [{\bf \bar y}_{i,1}, \dots, {\bf \bar y}_{i,K}]$, the obtained measurement matrix ${\bf \bar Y}_{i}$ can be expressed as
\begin{align}\label{yi}
	{\bf \bar Y}_{i} =  {\bf H}_{\text{R}} \bm{\Phi}_i {\bf H}_{\text{S}} \bm{\Psi}_i {\bf H}_{\text{T}} {\bf F} + {\bf \bar N}_{i} \in \mathbb{C}^{M_\text{R}\times K},
\end{align}
where ${\bf \bar N}_{i} \in \mathbb{C}^{M_\text{R}\times K}$ is expressed similarly. We assume that the training matrix ${\bf F}$ is designed with orthonormal  \kr{rows} so that ${\bf F} {\bf F}^{\mathsf{H}} = {\bf I}_{M_\text{T}}$, which directly implies that $K \geq M_\text{T}$. Then, after right filtering ${\bf \bar Y}_{i}$ with ${\bf F}^{\mathsf{H}}$, i.e., ${\bf Y}_{i} = {\bf \bar Y}_{i} {\bf F}^{\mathsf{H}}$, we can write the obtained matrix ${\bf Y}_{i}$ as  
\begin{align}\label{yi2}
	{\bf  Y}_{i} = {\bf H}_{\text{R}} \bm{\Phi}_i {\bf H}_{\text{S}} \bm{\Psi}_i {\bf H}_{\text{T}} + {\bf N}_{i} \in \mathbb{C}^{M_\text{R}\times M_\text{T}},
\end{align}
where ${\bf N}_{i} = {\bf \bar N}_{i} {\bf F}^{\mathsf{H}}$. Given the measurement matrices $	{\bf  Y}_{i}, \forall i \in\{1,\dots,I\}$, our main goal in Section \ref{sec2} is to obtain an estimate to the channel matrices ${\bf H}_{\text{R}}$, ${\bf H}_{\text{T}}$, and ${\bf H}_{\text{S}}$.

\vspace{-5pt}
\subsection{Proposed Tucker2-based for CE in D-RIS systems}\label{sec2}

By concatenating ${\bf  Y}_{1}, \dots, {\bf  Y}_{I} $  in (\ref{yi2}) behind each other, a 3-way tensor can be obtained as $\bm{\mathcal{Y}} =  [ {\bf  Y}_{1} ,  \sqcup_3  \dots, \sqcup_3  {\bf  Y}_I ]  \in \mathbb{C}^{M_\text{R} \times  M_\text{T} \times I}$, where ${\bf  Y}_{i}$ represents its $i$th frontal slice. Here, we note that the tensor $\bm{\mathcal{Y}}$ has a Tucker2 representation as \cite{andre_tensors}
\begin{align}\label{Ten}
	\bm{\mathcal{Y}} = \bm{\mathcal{S}} \times_1 {\bf H}_{\text{R}} \times_2 {\bf H}^{\mathsf{T}}_{\text{T}}  + \bm{\mathcal{N}}\in \mathbb{C}^{M_\text{R} \times  M_\text{T} \times I}, 
\end{align}
where $\bm{\mathcal{N}}$ is the noise tensor and $\bm{\mathcal{S}}$ is formed by concatenating $\bm{\Phi}_i {\bf H}_{\text{S}} \bm{\Psi}_i, \forall i\in\{1,\dots,I\}$, behind each other as 
\begin{align}
	\bm{\mathcal{S}} = [ \bm{\Phi}_1 {\bf H}_{\text{S}} \bm{\Psi}_1 ,  \sqcup_3  \dots, \sqcup_3  \bm{\Phi}_I {\bf H}_{\text{S}} \bm{\Psi}_I ]  \in \mathbb{C}^{N_\text{2} \times  N_\text{1} \times I}.
\end{align}

From the above, the CE problem can be formulated as
\begin{align}\label{Opt}
	\{{\bf \hat H}_{\text{R}}, {\bf \hat H}_{\text{T}}, {\bf \hat H}_{\text{S}}\} = \underset{ {\bf H}_{\text{R}}, {\bf H}_{\text{T}}, {\bf H}_{\text{S}} }{ \arg\min  } \Vert \bm{\mathcal{Y}} -  \bm{\mathcal{S}} \times_1 {\bf H}_{\text{R}} \times_2 {\bf H}^{\mathsf{T}}_{\text{T}} \Vert^{2}_{\text{F}},
\end{align}
which is nonconvex due to its joint optimization. To obtain a solution, we resort to an alternating minimization approach, where we solve (\ref{Opt}) for one variable assuming the other two are fixed. To achieve this end, we exploit the $n$-mode unfoldings of $\bm{\mathcal{Y}}$, i.e., $[\bm{\mathcal{Y}}]_{(n)}, n\in\{1,2,3\}$ expressed as  \cite{andre_tensors,Kolda}
\begin{align}
	[\bm{\mathcal{Y}}]_{(1)} &=  {\bf H}_{\text{R}} ~	{\bf Z}_\text{R}({{\bf H}_{\text{T}}},{{\bf H}_{\text{S}}} ) + [\bm{\mathcal{N}}]_{(1)}\in \mathbb{C}^{M_\text{R} \times  I M_\text{T} } \label{Y1}\\
	[\bm{\mathcal{Y}}]_{(2)} &=  {\bf H}^{\mathsf{T}}_{\text{T}} ~ {\bf Z}_\text{T}({{\bf H}_{\text{R}}},{{\bf H}_{\text{S}}} ) + [\bm{\mathcal{N}}]_{(2)} \in \mathbb{C}^{M_\text{T} \times  I M_\text{R} } \label{Y2} \\
	[\bm{\mathcal{Y}}]_{(3)} &=  [\bm{\mathcal{S}}]_{(3)}  ({\bf H}^{\mathsf{T}}_{\text{T}} \otimes {\bf H}_{\text{R}})^{\mathsf{T}} + [\bm{\mathcal{N}}]_{(3)} \in \mathbb{C}^{I \times  M_\text{T} M_\text{R} }, \label{Y3}
\end{align}  
where  ${\bf Z}_\text{R}({{\bf H}_{\text{T}}},{{\bf H}_{\text{S}}} ) = [\bm{\mathcal{S}}]_{(1)}  ({\bf I}_{I} \otimes {\bf H}^{\mathsf{T}}_{\text{T}})^{\mathsf{T}} \in \mathbb{C}^{ N_\text{2} \times  I M_\text{T} }$ and ${\bf Z}_\text{T}({{\bf H}_{\text{R}}},{{\bf H}_{\text{S}}} ) =   [\bm{\mathcal{S}}]_{(2)}  ({\bf I}_{I} \otimes {\bf H}_{\text{R}})^{\mathsf{T}} \in \mathbb{C}^{ N_\text{1} \times I M_\text{R} }$. Note that, \kr{according to the definition of $n$-mode unfoldings} \cite{Kolda}, $[\bm{\mathcal{S}}]_{(3)} \in \mathbb{C}^{I \times N_\text{1}  N_\text{2} } $ can be expressed as $[\bm{\mathcal{S}}]_{(3)} = (\bm{\Psi}  \diamond  \bm{\Phi}  )^{\mathsf{T}} \text{diag}\{ {\bf h}_{\text{S}}  \}$, 
where ${\bf h}_{\text{S}} = \text{vec}\{{\bf H}_{\text{S}}\}\in \mathbb{C}^{N_\text{1} N_\text{2}}$. Therefore, we have
\begin{align}\label{Y31}
	[\bm{\mathcal{Y}}]_{(3)} =  (\bm{\Psi}  \diamond  \bm{\Phi}  )^{\mathsf{T}} \text{diag}\{ {\bf h}_{\text{S}}  \}  ({\bf H}^{\mathsf{T}}_{\text{T}} \otimes {\bf H}_{\text{R}})^{\mathsf{T}} + [\bm{\mathcal{N}}]_{(3)}.
\end{align}

\begin{figure*}
	\centering
	\includegraphics[width=\figsize]{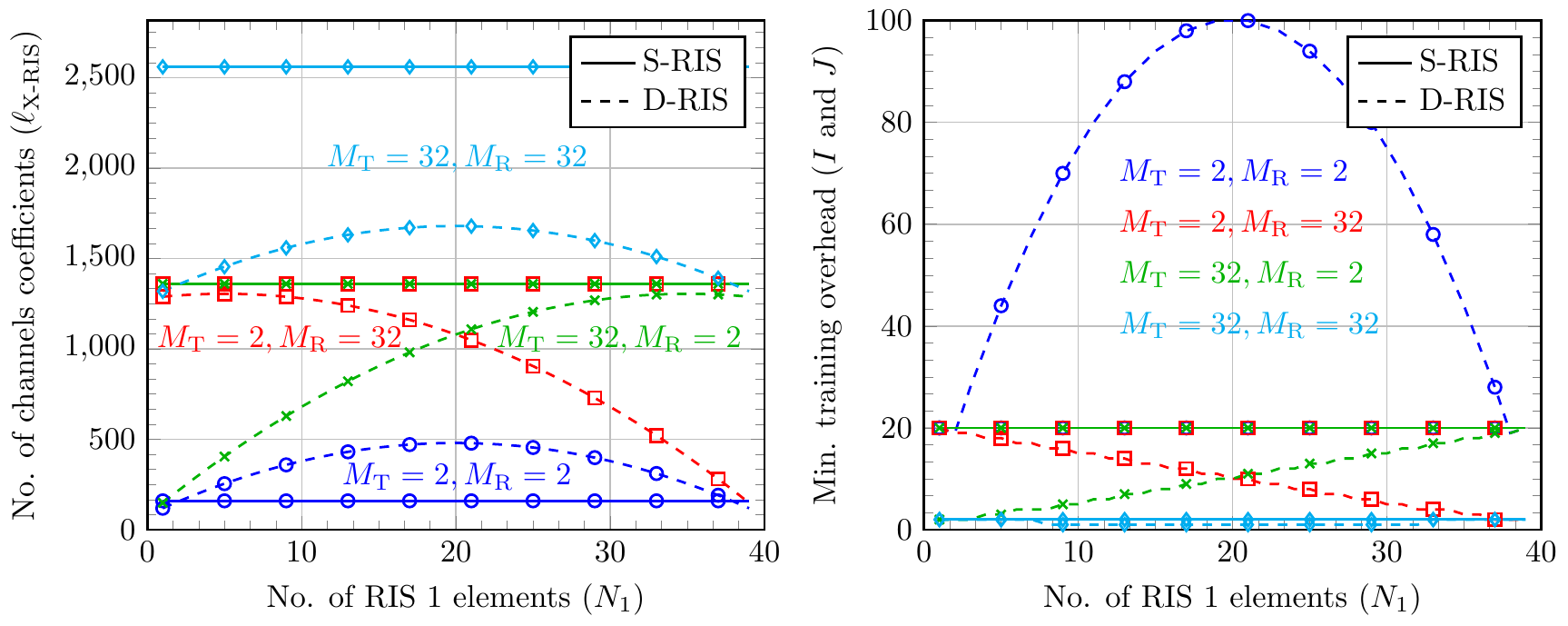}
	\vspace{-10pt}
	\caption{Number of channels coefficients ($\ell_{\text{X-RIS}}$) and minimum training overhead ($I$ and $J$) [$N = 40$, $N_2 = N - N_1$].}
	\vspace{-10pt}
	\label{fig:fig3}
\end{figure*}

The vectorized form of $[\bm{\mathcal{Y}}]_{(3)}$, i.e., ${\bf y}_{(3)} = \text{vec}\{[\bm{\mathcal{Y}}]_{(3)}\}$ can be expressed as 
\begin{align}\label{Yp1}
	{\bf y}_{(3)} = {\bf Z}_\text{S}({{\bf H}_{\text{T}}},{{\bf H}_{\text{R}}} ) ~ {\bf h}_{\text{S}} + {\bf n}_{(3)} \in \mathbb{C}^{I  M_\text{T} M_\text{R} } ,
\end{align}
where ${\bf n}_{(3)} = \text{vec}\{[\bm{\mathcal{N}}]_{(3)}\}$ , ${\bf Z}_\text{S}({{\bf H}_{\text{T}}},{{\bf H}_{\text{R}}} ) =\Big( ({\bf H}^{\mathsf{T}}_{\text{T}} \otimes {\bf H}_{\text{R}})  \diamond  (\bm{\Psi}  \diamond  \bm{\Phi}  )^{\mathsf{T}} \Big) \in \mathbb{C}^{ I M_\text{R} M_\text{T} \times  N_\text{1} N_\text{2}  }  $. By exploiting (\ref{Y1}), (\ref{Y2}), and (\ref{Yp1}), an estimate \kr{of} ${\bf H}_{\text{R}}$, ${\bf H}_{\text{T}}$, and ${\bf h}_{\text{S}}$ can be obtained as 
\begin{align}
	{\bf \hat H}_{\text{R}} &= \underset{ {\bf H}_{\text{R}} }{ \arg\min  } \Vert [\bm{\mathcal{Y}}]_{(1)} -  {\bf H}_{\text{R}} ~ {\bf Z}_\text{R}({{\bf H}_{\text{T}}},{{\bf H}_{\text{S}}} ) \Vert^{2}_{\text{F}} \label{pr}\\
	{\bf \hat H}_{\text{T}} &= \underset{ {\bf H}_{\text{T}} }{ \arg\min  } \Vert [\bm{\mathcal{Y}}]_{(2)} -  {\bf H}^{\mathsf{T}}_{\text{T}} ~	{\bf Z}_\text{T}({{\bf H}_{\text{R}}},{{\bf H}_{\text{S}}} ) \Vert^{2}_{\text{F}} \label{pt}\\
	{\bf \hat h}_{\text{S}} &= \underset{ {\bf h}_{\text{S}} }{ \arg\min  } \Vert {\bf y}_{(3)} -  {\bf Z}_\text{S}({{\bf H}_{\text{T}}},{{\bf H}_{\text{R}}} ) ~ {\bf h}_{\text{S}}  \Vert^{2}_{\text{2}}\label{ps}.
\end{align}

The above problems are convex and can be solved using the alternating least squares (ALS) method, as summarized in Algorithm \ref{ALS2RIS}, which is guaranteed to converge monotonically to, at least, a locally optimal solution \cite{andre_tensors}.

\begin{algorithm}
	\caption{ALS method for CE in D-RIS MIMO systems}
	\label{ALS2RIS}
	\begin{algorithmic}[1]
		\State{Input: Measurement tensor $\bm{\mathcal{Y}} \in \mathbb{C}^{M_\text{R} \times  M_\text{T} \times I} $ as in (\ref{Ten})}
		\State{Initialize: ${{\bf H}^{(0)}_{\text{T}}}$  and ${{\bf H}^{(0)}_{\text{S}}}$ and select $t_{\max}$. }
		\For{$t = 1$ to $t_{\max}$}
		
		\State{${{\bf H}^{(t)}_{\text{R}}} =  [\bm{\mathcal{Y}}]_{(1)} \{  {\bf Z}_\text{R}({{\bf H}^{(t-1)}_{\text{T}}},{{\bf H}^{(t-1)}_{\text{S}}} )  \}^{+}  $}
		
		\State{${\bf \hat H}^{(t)}_{\text{T}} = \big( [\bm{\mathcal{Y}}]_{(2)}  \{  {\bf Z}_\text{T}({{\bf H}^{(t)}_{\text{R}}},{{\bf H}^{(t-1)}_{\text{S}}} ) \}^{+} \big)^{\mathsf{T}}$}
		
		\State{${{\bf \hat H}^{(t)}_{\text{S}}} =  \text{unvec} \Big\{ \{{\bf Z}_\text{S}({{\bf H}^{(t)}_{\text{T}}},{{\bf H}^{(t)}_{\text{R}}} )\}^{+} {\bf y}_{(3)}  \Big\} $ }
		\EndFor
		
	\end{algorithmic}
\end{algorithm}

\vspace{-15pt}
\section{Comparison with S-RIS aided systems}
\vspace{-10pt}

In S-RIS-aided systems, on the other hand, the communication between the {Tx} and the {Rx} with $M_\text{T}$ and $M_\text{R}$ antennas, respectively, is aided via a single RIS with $N$ elements, as depicted \kr{on} the right-side of Fig. \ref{fig:fig1}. Let ${\bf G}_{\text{T}} \in \mathbb{C}^{N \times M_{\text{T}}}$ be the {Tx} to RIS channel and ${\bf G}_{\text{R}} \in \mathbb{C}^{M_{\text{R}}\times N}$ be the RIS to {Rx} channel. Then, it was shown in \cite{andreRIS,andre} that the received signals at the {Rx} can be arranged in a 3-way tensor admitting a canonical polyadic (CP) decomposition given as 
\begingroup\makeatletter\def\f@size{9.5}\check@mathfonts
\def\maketag@@@#1{\hbox{\m@th\small\normalfont#1}}% 
\begin{align}\label{Xten}
	\bm{\mathcal{X}} = \bm{\mathcal{I}}_{3,N} \times_1 {\bf G}_{\text{R}} \times_2 {\bf G}^{\mathsf{T}}_{\text{T}} \times_3 \bm{\Omega}^{\mathsf{T}} + \bm{\mathcal{E}} \in \mathbb{C}^{M_\text{R} \times  M_\text{T} \times J},  
\end{align} \endgroup
where $\bm{\mathcal{I}}_{3,N} \in \mathbb{Z}^{N \times N \times N}$ is the super-diagonal tensor, $\bm{\mathcal{E}}$ is the noise tensor, $\bm{\Omega} = [\bm{\omega}_1, \dots, \bm{\omega}_J] \in \mathbb{C}^{N \times J}$ is the RIS training matrix with $J$ training beams and $|[\bm{\omega}_j]_{[\ell]}| = 1/\sqrt{N}$. The $n$-mode unfoldings of $\bm{\mathcal{X}}$, $n\in\{1,2\}$, can be expressed as 
\begin{align}
	[\bm{\mathcal{X}}]_{(1)} &=  {\bf G}_{\text{R}}	~ {\bf V}_{\text{R}}({\bf G}_{\text{T}})  + [\bm{\mathcal{E}}]_{(1)}\in \mathbb{C}^{M_\text{R} \times  J M_\text{T} } \label{X1}\\
	[\bm{\mathcal{X}}]_{(2)} &=  {\bf G}^{\mathsf{T}}_{\text{T}} ~	{\bf V}_{\text{T}}({\bf G}_{\text{R}})   + [\bm{\mathcal{E}}]_{(2)} \in \mathbb{C}^{M_\text{T} \times  J M_\text{R} } \label{X2}, 
\end{align}
where ${\bf V}_{\text{R}}({\bf G}_{\text{T}}) = (\bm{\Omega}^{\mathsf{T}} \diamond {\bf G}^{\mathsf{T}}_{\text{T}})^{\mathsf{T}} \in \mathbb{C}^{ N \times  J M_\text{T} }$ and ${\bf V}_{\text{T}}({\bf G}_{\text{R}}) = (\bm{\Omega}^{\mathsf{T}} \diamond {\bf G}_{\text{R}})^{\mathsf{T}} \in \mathbb{C}^{ N \times  J M_\text{R} }$. Therefore, an ALS-based method, similarly to Algorithm \ref{ALS2RIS}, \kr{has been proposed} in \cite{andreRIS} to obtain an estimate \kr{of} ${\bf G}_{\text{R}}$ and ${\bf G}_{\text{T}}$, as summarized in Algorithm \ref{ALS1RIS}. 
\begin{algorithm}
	\caption{ALS method for CE in S-RIS MIMO systems}
	\label{ALS1RIS}
	\begin{algorithmic}[1]
		\State{Input: Measurement tensor $\bm{\mathcal{X}} \in \mathbb{C}^{M_\text{R} \times  M_\text{T} \times J} $ as in (\ref{Xten})}
		\State{Initialize: ${{\bf G}^{(0)}_{\text{T}}}$ and select $t_{\max}$. }
		\For{$t = 1$ to $t_{\max}$}
		\State{${{\bf G}^{(t)}_{\text{R}}} = [\bm{\mathcal{X}}]_{(1)}  \{{\bf V}_{\text{R}}({\bf G}^{(t-1)}_{\text{T}})\}^{+} $  }
		\State{${{\bf G}^{(t)}_{\text{T}}} = \big([\bm{\mathcal{X}}]_{(2)}  \{{\bf V}_{\text{T}}({\bf G}^{(t)}_{\text{R}})\}^{+}  \big)^{\mathsf{T}} $  }
		\EndFor
	\end{algorithmic}
\end{algorithm} 
\begin{figure*}
	\centering
	\includegraphics[width=\figsize]{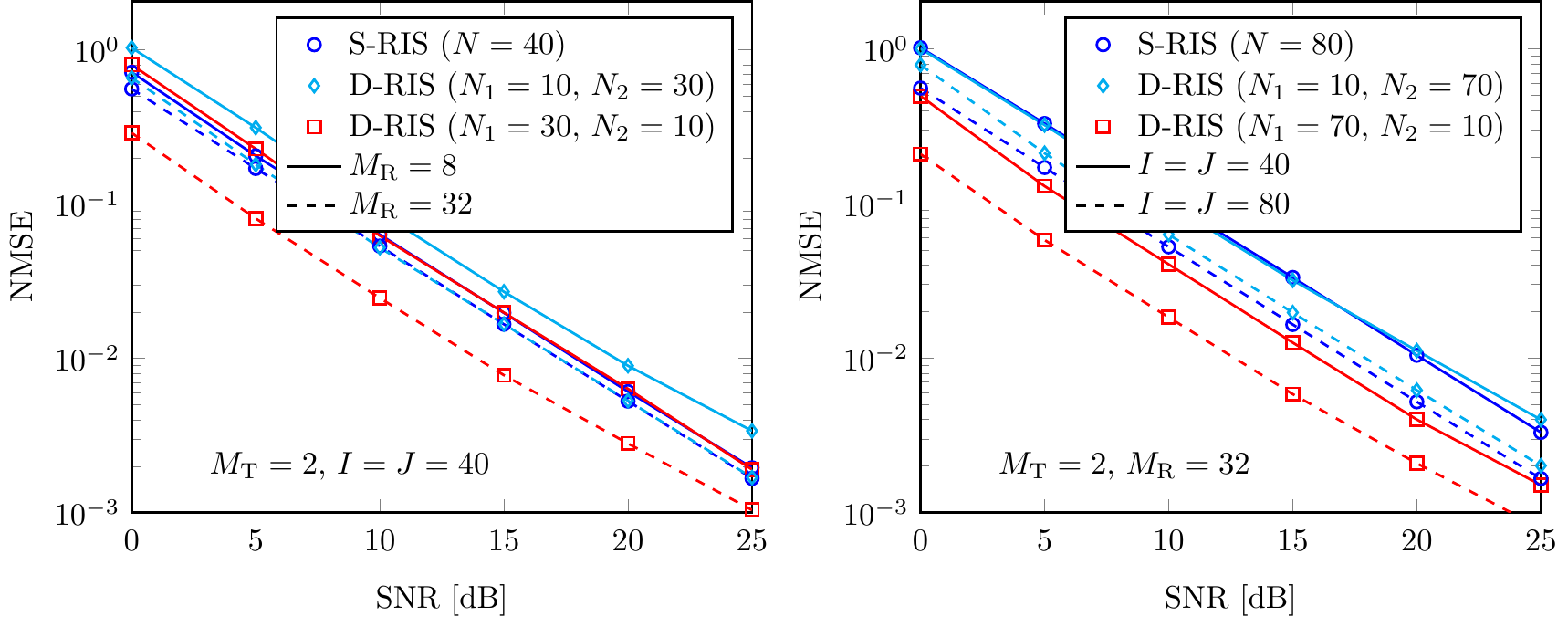}
	\vspace{-12pt}
	\caption{{NMSE versus SNR comparing the D-RIS against the S-RIS systems for different system settings [$M_\text{T} = 2$].}}
	\vspace{-10pt}
	\label{fig:fig4}
\end{figure*}

\textbf{Identifiablity}, in the LS sense, can be obtained by noting that ${\bf Z}_\text{R}$, ${\bf Z}_\text{T}$, ${\bf V}_\text{R}$, and ${\bf V}_\text{T}$ need to \kr{have} full column-rank, while ${\bf Z}_\text{S}$ needs to \kr{have} full row-rank \cite{andre_tensors}. This leads to the following conditions: $I M_\text{T} \geq N_\text{2}$ (for ${\bf Z}_\text{R}$), $I M_\text{R} \geq N_\text{1}$ (for ${\bf Z}_\text{T}$), $J M_\text{T} \geq N$ (for ${\bf V}_\text{R}$), $J M_\text{R} \geq N$ (for ${\bf V}_\text{T}$), and $I M_\text{R} M_\text{T} \geq N_\text{1} N_\text{2}$ (for ${\bf Z}_\text{S}$). Therefore, we conclude that  
\begingroup\makeatletter\def\f@size{9.5}\check@mathfonts
\def\maketag@@@#1{\hbox{\m@th\small\normalfont#1}}%  
\begin{align}
	I &\geq \max\big\{ \ceil*{{N_\text{2}}/{M_\text{T}}}, ~~\ceil*{{N_\text{1}}/{M_\text{R}}}, ~~ \ceil*{{N_\text{1} N_\text{2}}/{M_\text{R} M_\text{T}}} \big\} \label{dris} \\
	J &\geq \max\big\{ \ceil*{ {N}/{M_\text{T}} }, ~~ \ceil*{{N}/{M_\text{R}}} \big\}, \label{sris}
\end{align}\endgroup
\ka{where (\ref{dris}) is for the D-RIS systems and (\ref{sris}) is for the S-RIS systems.}  
Let $\ell_{\text{D-RIS}}$ and $\ell_{\text{S-RIS}}$ denote the total number of channel coefficients in the D-RIS and the S-RIS communication scenarios, respectively, which are given as 
\begin{align}
	\ell_{\text{D-RIS}}  &= M_\text{T} N_1 + N_1 N_2 + M_\text{R} N_2 \label{ldris} \\
	\ell_{\text{S-RIS}}  &= M_\text{T} N + M_\text{R} N \label{lsris}.
\end{align} 

Let us assume that the S-RIS elements $N$ are distributed between RIS 1 and RIS 2 in the D-RIS scenario such that $N = N_1 + N_2$. In Fig. \ref{fig:fig3}  we plot results of (\ref{dris}), (\ref{sris}), (\ref{ldris}), and (\ref{lsris}) for different $M_\text{R}$ and $M_\text{T}$ values assuming $N = 40$. Note that \kr{along} the $x$-axis we vary $N_1$ so that $N_2 = N - N_1$. From Fig. \ref{fig:fig3}, we have the following remarks:

\textit{\textbf{Remark 1}: If $N \gg \max \{{M_\text{R}},{M_\text{T}}\}$ and ${M_\text{R}} \approx {M_\text{T}}$, then S-RIS, i.e., Algorithm \ref{ALS1RIS} requires less training overhead compared to D-RIS, i.e., Algorithm \ref{ALS2RIS}, in most of the $N_1$ and $N_2$ distribution scenarios. This comes from the fact that the number of channel coefficients that D-RIS needs to estimate, i.e., $ \ell_{\text{D-RIS}}$ is much larger than that of S-RIS, i.e., $\ell_{\text{S-RIS}}$. }  

\textit{\textbf{Remark 2}: If $N \approx \max \{{M_\text{R}},{M_\text{T}}\}$, then D-RIS requires less training overhead compared to S-RIS for the same reason mentioned in Remark 1, i.e., $\ell_{\text{D-RIS}} < \ell_{\text{S-RIS}}$.}

\textit{\textbf{Remark 3}: \ka{In the D-RIS systems, the careful distribution of the $N$ elements between RIS 1 and RIS 2 (i.e., $N_1$ and $N_2$) can reduce the training overhead of Algorithm \ref{ALS2RIS}. From Fig.~\ref{fig:fig3}, we can note that the best distribution depends on the ${M_\text{R}}$ and the $ {M_\text{T}}$ values as: if ${M_\text{R}} > {M_\text{T}}$, then it is more beneficial to allocate more elements to RIS 1 than RIS~2, i.e., $N_1 > N_2$. This observation is \kr{reversed} if ${M_\text{R}} < {M_\text{T}}$, i.e., more elements should be allocated to RIS 2 than RIS~1 as $N_1 < N_2$. }}

%\textit{\textbf{Conjecture:} Since the system spectral efficiency is inversely proportional to the length of the training overhead, we can conjecture that there is an optimal distribution of the $N$ elements between RIS 1 and RIS 2 that strikes an optimal trade-off between the training overhead and the achievable performance, which we leave for future work. } 

\textbf{Computational complexity}: Assuming that \kr{the conditions in} (\ref{dris}) and (\ref{sris}) are satisfied, the \ka{complexities}\footnote{Here, we have assumed that the complexity of calculating the Moore-Penrose inverse of a $n\times m$ matrix is on the order of $\mathcal{O}(\min\{n,m\}^3)$.} of Algorithm \ref{ALS2RIS} and Algorithm \ref{ALS1RIS} are on the order of $\mathcal{O} \big( t_{\max} \cdot ( N^3_2 + N^3_1 + (N_1 \cdot N_2)^3)\big)$ and $\mathcal{O} \big(t_{\max} \cdot 2N^3\big)$, respectively.

\textbf{Ambiguities}: Assuming that \kr{the conditions in} (\ref{dris}) and (\ref{sris}) are satisfied, then the estimated MIMO channels by Algorithm~\ref{ALS2RIS} and Algorithm~\ref{ALS1RIS} are unique up to scalar ambiguities \cite{andreRIS,andre_tensors}. In particular, the estimated channels are related to the perfect (true) channels as: \kr{${\bf \hat H}_{\text{R}} \approx  {\bf H}_{\text{R}} \bm{\Delta}_{\text{R}} $, ${\bf \hat H}_{\text{T}} \approx \bm{\Delta}_{\text{T}} {\bf H}_{\text{T}}$, ${\bf H}_{\text{S}} \approx \bm{\Delta}^{-1}_{\text{R}} {\bf \hat H}_{\text{S}} \bm{\Delta}^{-1}_{\text{T}}$, ${\bf \hat G}_{\text{R}} \approx {\bf G}_{\text{R}} \bm{\Lambda}$, and ${\bf \hat G}_{\text{T}} \approx \bm{\Lambda}^{-1} {\bf \hat G}_{\text{T}}$}, where $\bm{\Delta}$ and $\bm{\Lambda}$ are diagonal matrices holding the scaling ambiguities. However, these ambiguities disappear when reconstructing an estimate of the effective end-to-end channels ${\bf \hat H}_{\text{e}} = {\bf \hat H}_{\text{R}} {\bf \hat H}_{\text{S}} {\bf \hat H}_{\text{T}}$ and ${\bf \hat G}_{\text{e}} = {\bf \hat G}_{\text{R}} {\bf \hat G}_{\text{T}}$. Moreover, note that, due to the knowledge of the RIS reflection matrices $\bm{\Psi}$, $\bm{\Phi}$, and $\bm{\Omega}$ at the {Rx}, the permutation ambiguities do not exist \cite{andreRIS}.

\vspace{-5pt}
\section{Simulation Results}
\vspace{-5pt}
We assume that the entries of \kr{the channel matrices} ${\bf H}_{\text{R}}$, ${\bf H}_{\text{T}}$, ${\bf H}_{\text{S}}$, ${\bf G}_{\text{R}}$, and $ {\bf G}_{\text{T}}$
\kr{follow a Rayleigh fading distribution}. We show results in terms of \kr{the} normalized-mean-square-error (NMSE) of \kr{the} effective channels defined as $\text{NMSE} = \mathbb{E} \big\{{\Vert {\bf H}_{\text{e}} - {{\bf \hat H}}_{\text{e}} \Vert^{2}_{\text{F}} \big\} }/\mathbb{E} \big\{{\Vert {\bf H}_{\text{e}}\Vert^{2}_{\text{F}}} \big\}$, for \kr{the} D-RIS, and $\text{NMSE} = \mathbb{E} \big\{{\Vert {\bf G}_{\text{e}} - {{\bf \hat G}}_{\text{e}} \Vert^{2}_{\text{F}} \big\} }/\mathbb{E} \big\{{\Vert {\bf G}_{\text{e}}\Vert^{2}_{\text{F}}} \big\}$, for \kr{the} S-RIS. We define the signal-to-noise ratio (SNR) as $\text{SNR} = \mathbb{E} \big\{{\Vert \bm{\mathcal{Y}} - \bm{\mathcal{N}}\Vert^{2}_{\text{F}}  }\big\}/\mathbb{E}\big\{{\Vert \bm{\mathcal{N}}\Vert^{2}_{\text{F}}} \big\}$, for \kr{the} D-RIS, and $\text{SNR} = \mathbb{E} \big\{{\Vert \bm{\mathcal{X}} - \bm{\mathcal{E}}\Vert^{2}_{\text{F}}  }\big\}/\mathbb{E}\big\{{\Vert \bm{\mathcal{E}}\Vert^{2}_{\text{F}}} \big\}$, for \kr{the} S-RIS. Moreover, assuming that $ I \leq N_1 N_2$ and $J \leq N$, the training matrices ${\bf \Psi}$,  ${\bf \Phi}$, and ${\bm \Omega}$ are updated using a DFT-based approach as: ${\bf \Phi}  = [{\bf W}_{N_2} \otimes \bm{1}^{\mathsf{T}}_{ \bar{I}_2 }]_{[:,1:I]}$, ${\bf \Psi}  = [\bm{1}^{\mathsf{T}}_{ \bar{I}_1 } \otimes {\bf W}_{N_1}]_{[:,1:I]}  $, and ${\bm \Omega}  = [{\bf W}_{N}]_{[:, 1:J]} $, where $\bar{I}_1 = \ceil*{ \frac{I}{N_1} }$, $\bar{I}_2 = \ceil*{ \frac{I}{N_2} }$, and ${\bf W}_{K}$ is the normalized $K\times K$ DFT matrix such that ${\bm \Omega}^{\mathsf{H}} {\bm \Omega} = {\bf I}_{ J }$ and $\bm{\Upsilon}^{\mathsf{H}} \bm{\Upsilon} = {\bf I}_{ I }$, where $\bm{\Upsilon} \eqbydef \bm{\Psi}  \diamond  \bm{\Phi}$.

Fig. \ref{fig:fig4} shows the NMSE versus SNR results for different system settings. From the left-side figure, we can see that when $M_\text{R} = 8$, \kr{the} D-RIS, i.e., Algorithm~\ref{ALS2RIS} has a worse NMSE performance compared to the S-RIS, i.e., Algorithm~\ref{ALS1RIS} especially with the $[N_1,N_2] = [10,30]$ distribution scenario. This can be explained from Fig. \ref{fig:fig3} and Remarks~1~and~3. Note that in a such system setting, \kr{the} D-RIS has a larger number of channels coefficients $\ell_{\text{D-RIS}} = 560$ compared to \kr{the} S-RIS $\ell_{\text{S-RIS}} = 400$. Moreover, as we have highlighted in Remark 3, we can see that \kr{the} $[N_1,N_2] = [30,10]$ distribution scenario has a better NMSE performance than $[N_1,N_2] = [10,30]$, since $M_\text{T}  < M_\text{R}$. On the  other hand, when $M_\text{R} = 32$, we can see that \kr{the} D-RIS has a much better NMSE performance compared to \kr{the} S-RIS, especially with the $[N_1,N_2] = [30,10]$ distribution scenario. This can be explained \kr{in the same way from} Fig. \ref{fig:fig3} and Remarks~2 and~3. From the right-side figure, we can see that the same observations hold true when we increase $N$ from $40$ to $80$ or when we increase the training overhead $I$ and $J$ from $40$ to $80$.

\vspace{-10pt}
\section{Conclusions}
\vspace{-10pt}
In this paper, we have shown that D-RIS MIMO systems can be used to reduce the training overhead and to improve the channel estimation accuracy compared to S-RIS aided systems. This comes from the observation that if the RIS elements in the S-RIS system are distributed carefully between the two RISs in the D-RIS system, the number of channel coefficients in the D-RIS system that need to be estimated reduces significantly compared to the S-RIS system. Therefore, D-RIS systems can be seen as an appealing approach to further increase the coverage, capacity, and efficiency of wireless networks compared to S-RIS systems.

\vfill\pagebreak
%\bibliographystyle{IEEEbib}
%\bibliography{strings,refs}

\bibliographystyle{IEEEtran}
\bibliography{refs}	

\begin{thebibliography}{10}
\providecommand{\url}[1]{#1}
\csname url@samestyle\endcsname
\providecommand{\newblock}{\relax}
\providecommand{\bibinfo}[2]{#2}
\providecommand{\BIBentrySTDinterwordspacing}{\spaceskip=0pt\relax}
\providecommand{\BIBentryALTinterwordstretchfactor}{4}
\providecommand{\BIBentryALTinterwordspacing}{\spaceskip=\fontdimen2\font plus
\BIBentryALTinterwordstretchfactor\fontdimen3\font minus
  \fontdimen4\font\relax}
\providecommand{\BIBforeignlanguage}[2]{{%
\expandafter\ifx\csname l@#1\endcsname\relax
\typeout{** WARNING: IEEEtran.bst: No hyphenation pattern has been}%
\typeout{** loaded for the language `#1'. Using the pattern for}%
\typeout{** the default language instead.}%
\else
\language=\csname l@#1\endcsname
\fi
#2}}
\providecommand{\BIBdecl}{\relax}
\BIBdecl

\bibitem{comMagazine}
C.~{Liaskos}, S.~{Nie}, A.~{Tsioliaridou}, A.~{Pitsillides}, S.~{Ioannidis},
  and I.~{Akyildiz}, ``A new wireless communication paradigm through
  software-controlled metasurfaces,'' \emph{IEEE Commun. Mag.}, vol.~56, no.~9,
  pp. 162--169, 2018.

\bibitem{irs}
M.~Di~Renzo, A.~Zappone, M.~Debbah, M.-S. Alouini, C.~Yuen, J.~de~Rosny, and
  S.~Tretyakov, ``Smart radio environments empowered by reconfigurable
  intelligent surfaces: How it works, state of research, and the road ahead,''
  \emph{IEEE J. Sel. Areas Commun.}, vol.~38, no.~11, pp. 2450--2525, 2020.

\bibitem{RISCapacity}
S.~Zhang and R.~Zhang, ``Capacity characterization for intelligent reflecting
  surface aided {MIMO} communication,'' \emph{IEEE J. Sel. Areas Commun.},
  vol.~38, no.~8, pp. 1823--1838, Aug. 2020.

\bibitem{ardah2020trice}
K.~Ardah, S.~Gherekhloo, A.~L.~F. de~Almeida, and M.~Haardt, ``{TRICE}: A
  channel estimation framework for {RIS}-aided millimeter-wave {MIMO}
  systems,'' \emph{IEEE Signal Process. Lett.}, vol.~28, pp. 513--517, Feb.
  2021.

\bibitem{gherekhloo2021tensor}
S.~Gherekhloo, K.~Ardah, A.~L. de~Almeida, and M.~Haardt, ``Tensor-based
  channel estimation and reflection design for {RIS}-aided millimeter-wave
  {MIMO} communication systems,'' \emph{arXiv preprint arXiv:2107.13851}, 2021.

\bibitem{BF1}
Q.~{Wu} and R.~{Zhang}, ``Intelligent reflecting surface enhanced wireless
  network via joint active and passive beamforming,'' \emph{IEEE Trans.
  Wireless Commun.}, vol.~18, no.~11, pp. 5394--5409, 2019.

\bibitem{andre}
G.~T. {de Araújo} and A.~L.~F. {de Almeida}, ``{PARAFAC}-based channel
  estimation for intelligent reflective surface assisted {MIMO} system,'' in
  \emph{Proc. IEEE 11th Sensor Array and Multichannel Signal Processing
  Workshop (SAM)}, 2020, pp. 1--5.

\bibitem{andreRIS}
G.~T. de~Araújo, A.~L.~F. de~Almeida, and R.~Boyer, ``Channel estimation for
  intelligent reflecting surface assisted {MIMO} systems: A tensor modeling
  approach,'' \emph{IEEE J. Sel. Topics Signal Process.}, vol.~15, no.~3, pp.
  789--802, 2021.

\bibitem{Emil}
E.~Björnson and L.~Sanguinetti, ``Power scaling laws and near-field behaviors
  of massive {MIMO} and intelligent reflecting surfaces,'' \emph{IEEE Open J.
  Commun. Soc.}, vol.~1, pp. 1306--1324, 2020.

\bibitem{Kolda}
T.~G. Kolda and B.~W. Bader, ``Tensor decompositions and applications,''
  \emph{SIAM Review}, vol.~51, no.~3, pp. 455--500, Sept. 2009.

\bibitem{andre_tensors}
P.~Comon, X.~Luciani, and A.~L.~F. de~Almeida, ``Tensor decompositions,
  alternating least squares and other tales,'' \emph{Journal of Chemometrics},
  vol.~23, no. 7‐8, pp. 393--405, 2009.

\bibitem{ardah2019wsa}
K.~{Ardah}, A.~L.~F. de~{Almeida}, and M.~{Haardt}, ``Low-complexity millimeter
  wave {CSI} estimation in {MIMO-OFDM} hybrid beamforming systems,'' in
  \emph{Proc. 23rd International ITG Workshop on Smart Antennas (WSA)}, Apr.
  2019, pp. 1--5.

\bibitem{CoLocated}
V.~Jamali, A.~M. Tulino, G.~Fischer, R.~R. Müller, and R.~Schober,
  ``Intelligent surface-aided transmitter architectures for millimeter-wave
  ultra massive {MIMO} systems,'' \emph{IEEE Open J. Commun. Soc.}, vol.~2, pp.
  144--167, 2021.

\end{thebibliography}

\end{document}